\documentclass{llncs}
\usepackage{llncsdoc}
\usepackage{tabularx,colortbl}
\usepackage[table]{xcolor}
\usepackage[pdftex]{graphicx}
\usepackage{subfigure}
\usepackage{amssymb}
\usepackage{mathrsfs}
\usepackage{appendix}

\newcommand{\delete}[1]{{}}

\newenvironment{keywords}{
       \list{}{\advance\topsep by0.35cm\relax\small
       \leftmargin=1cm
       \labelwidth=0.35cm
       \listparindent=0.35cm
       \itemindent\listparindent
       \rightmargin\leftmargin}\item[\hskip\labelsep
                                     \bfseries Keywords:]}
     {\endlist}

\begin{document}

\title{Game Characterizations of Timed Relations for Timed Automata Processes}

\author{Shibashis Guha  $ ^1$ and $\:$
Shankara Narayanan Krishna $ ^2$
}

\institute{$ ^1$Department of Computer Science and Engineering,\\ Indian Institute of Technology Delhi\\
\email{shibashis@cse.iitd.ac.in}\\[.2cm]
$ ^2$Department of Computer Science and Engineering,\\ Indian Institute of Technology Bombay\\
\email{krishnas@cse.iitb.ac.in}}
\maketitle

\begin{abstract}
In this work, we design the game semantics for timed equivalences and preorders of timed processes. 
The timed games corresponding to the various timed relations form a hierarchy. These games are similar to Stirling's bisimulation games. If it is the case that the existence of a winning strategy for the defender in a game ${\cal G}_1$ implies that there exists a winning strategy for the defender in another game ${\cal G}_2$, then the relation that corresponds to ${\cal G}_1$ is stronger than the relation corresponding to ${\cal G}_2$. The game hierarchy also throws light into several  timed relations that are not considered in this paper.
\end{abstract}
\begin{keywords}
Timed automata, bisimulation, timed transition system, timed games, EF games
\end{keywords}

\section{Introduction}\label{sec-intro}
\begin{figure}
\centering
\includegraphics[width=0.9\textwidth]{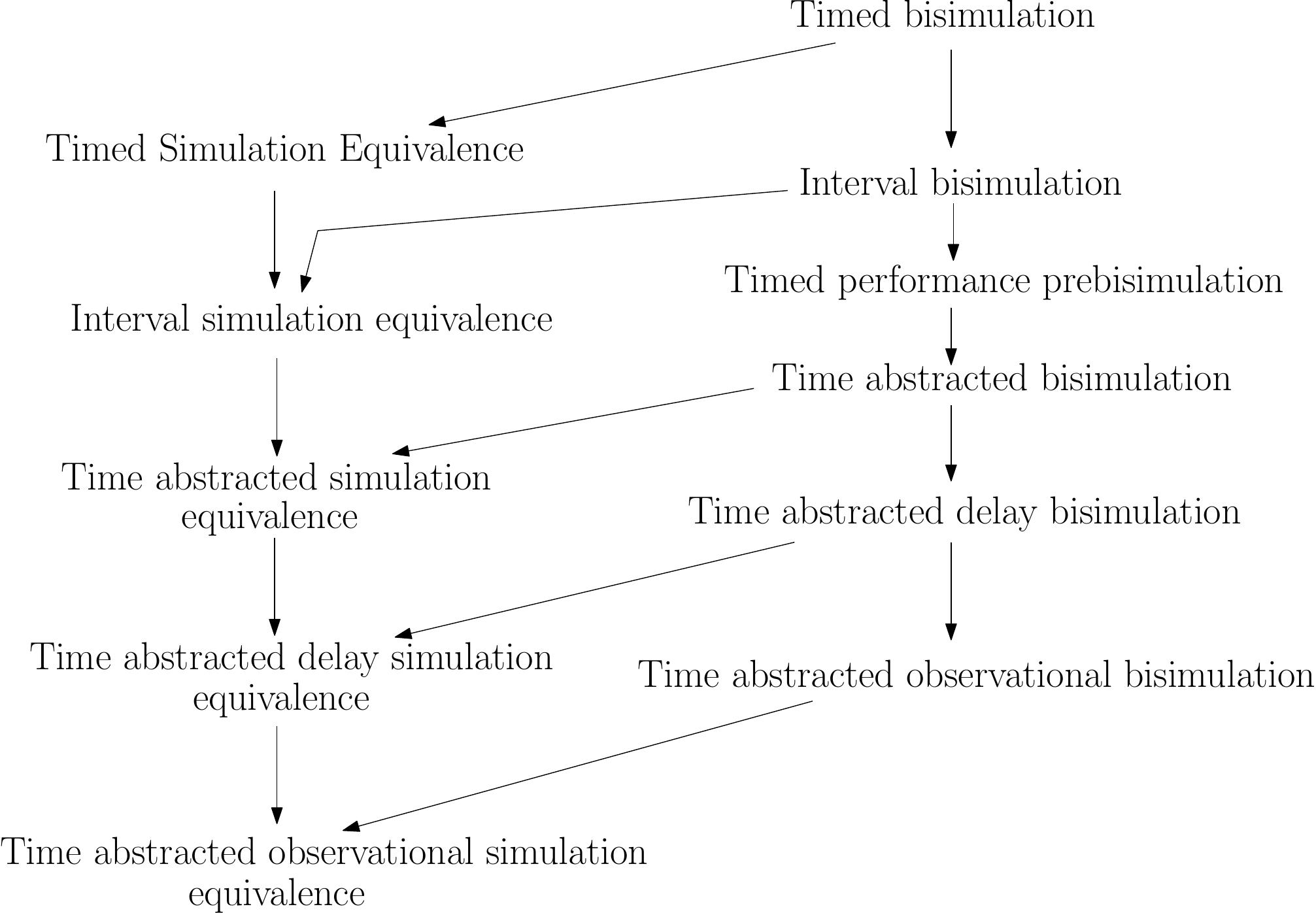}
\caption{\label{fig-timedSpectrum} Spectrum of timed relations}
\end{figure}

Bisimulation games \cite{CS1} have been defined for discrete procsses. Surpisingly,  there are no game semantics for similar relations with real time. In this paper, we extend bisimulation games to provide a coherent game structure for equivalence and preorder relations that involve real time. In \cite{VG1}, several semantic equivalences have been defined and compared in a model independent way. Some of these equivalences have been extended for real time as well. For example, there are well known notions of equivalences which include timed bisimulation and time abstracted bisimulation. In \cite{TY1}, equivalences even weaker than time abstracted bisimulation have been defined. They are time abstracted delay bisimulation and time abstracted observational bisimulation. In timed bisimulation, every time delay needs to be matched exactly which makes it a very strong form of equivalence. Time abstracted bisimulation on the other hand is a much weaker form of equivalence where a time delay by one process can be matched by any delay so that the respective derivatives are time abstracted bisimilar. To bridge this gap, in this paper we introduce \emph{interval bisimulation} which lies in between timed and time abstracted bisimulation. We can also conceive of a simulation relation corresponding to each of these bisimulation relations. In this work, we consider the hierarchy of these timed relations. Apart from proposing interval bisimulation and simulation equivalences corresponding to each well known bisimulation relation, the main contribution of this work includes proposing a generalized game semantics for these timed relations. This generalized game semantics will have certain parameters which being assigned different values can correspond to each of the relations shown in figure \ref{fig-timedSpectrum}. For the sake of completion of this spectrum of timed relations, we also include \emph{timed performance prebisimulation} which has been proposed in the recent work \cite{GNA1}. In this work, more particularly, we propose the game semantics for \emph{timed automata processes}. We choose timed automata since it is a well studied formalism and the decidability results for many of the relations based on timed automata are known. In contrast to Van Glabbeek's spectrum, at this point of time, we do not consider any form of trace equivalence in our work. We also do not consider either timed counterparts of relations like 2-nested simulation preorder or ready equivalence since the study of such relations are not known with respect to real time to the best of our knowledge. Game semantics for the equivalences in Van Glabbeek's spectrum  has been proposed in \cite{CD1}. In figure \ref{fig-timedSpectrum}, we present a spectrum of the timed relations mentioned above. In section \ref{sec-ta}, we present timed automata and its semantics briefly. In section \ref{sec-zone}, we present zone valuation graph as defined in \cite{GNA1}. Section \ref{sec-timedrels} describes several timed relations and compares them.   In section \ref{sec-timedgames}, we present the game characterizations of these timed relations. In section \ref{sec-gamehier}, we provide several lemmas that can be used to construct the hierarchy of the timed games. We conclude in section \ref{sec-conc}.

\section{Timed Automata}\label{sec-ta}
\emph{Timed automata}~\cite{AD1} is an approach to model time critical systems where the system is modeled with \emph{clocks} that track elapsed time. Timing of actions and time invariants on states can be specified using this model.

A timed automaton is a finite-state structure which can manipulate real-valued clock variables. Corresponding to every transition, a subset of the clocks can be specified that can be \emph{reset} to zero. In this paper, the clocks that are reset in a transition are shown as being enclosed in braces. Clock constraints also specify the condition for actions being enabled. If the constraints are not satisfied, the actions will be disabled. Constraints can also be used to specify the amount of time that may be spent in a location. The clock constraints $\mathcal{B}(C)$ over a set of clocks $C$ is given by the following grammar:
\begin{center}
$g ::= \; x \smile c \:|\: g \wedge g$
\end{center}
where $c \in \mathbb{N}$ and $x \in C$ and $\smile \: \in \: \{\le, <, =, >, \ge\}$.
A \emph{timed automaton} over a finite set of clocks $C$ and a finite set of actions $Act$ is a quadruple $(L,l_0, E, I)$ \cite{LAKJ1} where
$L$ is a finite set of locations, ranged over by $l$,
$l_0 \in L$ is the initial location,
$E \subseteq L \:\times\: \mathcal{B}(C) \:\times\: Act \:\times\: 2^C \:\times\: L$ is a finite set of \emph{edges}, and 
$I\: : \: L \rightarrow \mathcal{B}(C)$ assigns invariants to locations.
\subsection{Semantics} The semantics of a timed automaton can be described with a \emph{timed labeled transition system}(TLTS)\cite{LAKJ1}. Let $A= \: (L, l_0, E, I)$ be a timed automaton over a set of clocks $C$ and a set of visible actions $Act$. The timed transition system $T(A)$ generated by $A$ can be defined as $T(A) = (Proc, Lab, \{\stackrel{\alpha}{\longrightarrow} | \alpha \in Lab\})$, where $Proc \: = \: \{(l,v)\:|\: (l,v) \in L \: \times \: (C \rightarrow \mathbb{R}_{\ge 0})$ and $v \models I(l)\}$, i.e. \emph{states}
are of the form $(l,v)$, where $l$ is a location of the timed automaton and $v$ is a valuation that satisfies the invariant of $l$. We use the terms \emph{process} and \emph{state} interchangeably in this text.
$Lab = Act \cup \mathbb{R}_{\ge 0}$ is the set of labels; and
the transition relation is defined by 
$(l,v) \stackrel{a}{\longrightarrow} (l', v')$ if for an edge $(l \stackrel {g,a,r}{\longrightarrow}l') \in \: E$, $v \models g, v' = v[r]$ and $v' \models I(l')$, where an edge $(l \stackrel {g,a,r}{\longrightarrow}l')$ denotes that $l$ is the source location, $g$ is the guard, $a$ is the action, $r$ is the set of clocks to be reset and $l'$ is the target location. $(l,v)\stackrel{d}{\longrightarrow}(l, v+d)$ for all $d \in \mathbb{R}_{\ge 0}$ such that $v \models I(l)$ and $v+d \models I(l)$ where $v + d$ is the valuation in which every clock value is incremented by $d$.
Let $v_0$ denote the valuation such that $v_0(x)=0$ for all $x \in C$. If $v_0$ satisfies the invariant condition of the initial location $l_0$, then $(l_0, v_0)$ is the initial state or the initial configuration of $T(A)$.

\section{Graph Structure for Games}\label{sec-zone}
A bisimulation game \cite{CS1}\cite{CD1} is a two player game and consists of two \emph{graph structures} on which the game is played. The graphs are the visual representation of the two process descriptions for which the existence of a bisimulation relation has to be checked. For the games corresponding to timed equivalence and timed preorder relations too, we need to use a graph structure on which such timed games can be played. In this paper, we show how zone valuation graph \cite{GNA1} and some of its \emph{variants} are used as the graph structure on which the games corresponding to the timed relations are played. One must note that zone valuation graph \emph{cannot} be directly used in all the games discussed later. We may require certain modifications in the graph structure to characterize the games for various timed relations.

We briefly describe zone valuation graph. For this we first introduce zone and zone graph. The following two definitions are from \cite{WL1}.

\subsection{Zone Valuation Graph} \label{zonevalgraph}
\begin{definition} \textbf{zone:}
The characteristic set of a linear formula $\phi$, a clock constraint of the form $x \smile c$ or a diagonal constraint of the form $x - y \smile c$, where $x, y \in C$, is the set of all valuations for which $\phi$ holds. A zone is a finite union of characteristic sets.
\end{definition}

A \emph{zone graph} is similar to a \emph{region graph}\cite{LAKJ1} with the difference that each node consists of a timed automaton location and a zone.
\begin{definition}\textbf{zone graph:}
For a timed automaton $P = (L, l_0, E, I)$, a zone graph is a transition system $(S, s_0, Lep, \rightarrow)$, where $Lep = Act \cup \{\varepsilon\}$, 
  $\varepsilon$ is an action corresponding to delay transitions of the processes of the zone, $S \subseteq L \times \Phi_{\vee}(C)$ is the set of nodes, $s_0 = (l_0, \phi_0(C))$, $\rightarrow \subseteq S \times Lep \times S$ is connected, $\phi_0(C)$ is the formula where all the clocks in $C$ are 0 and $\Phi_{\vee}(C)$ denotes the set of all zones.
\end{definition}
\begin{definition}\textbf{Bisimulation between zone graphs}
\item For two zone graphs, $Z_{1} = (S_1, s_1, Lep, \rightarrow_1)$ and $Z_{2} = (S_2,$ $s_2,$ $Lep,$ $\rightarrow_2)$, $Z_{1}$ is strongly bisimilar\cite{RM1} to $Z_{2}$, denoted as $Z_{1} \sim Z_{2}$, iff the nodes $s_1$ and $s_2$ are strongly bisimilar, denoted by $s_1 \sim s_2$.
\end{definition}
While checking strong bisimulation between the two zone graphs, $\varepsilon$ is considered visible similar to an action in $Act$. The $\varepsilon$ action represents a process delay $d \in \mathbb{R}_{\ge 0}$, where $d \ge 0$. Hence each node in the zone valuation graph has an $\varepsilon$ transition to itself. Besides as in region graph, $\varepsilon$ transitions are transitive in nature. To avoid clutter, the self loops and the transitive $\varepsilon$ transitions are not shown in any of the zone valuation graphs in this paper.
We present here zone valuation graph as defined in \cite{GNA1}. One should note that a zone valuation graph corresponds to a particular timed process or valuation of the timed automaton.

\begin{figure}
\centering
\includegraphics[width=0.9\textwidth]{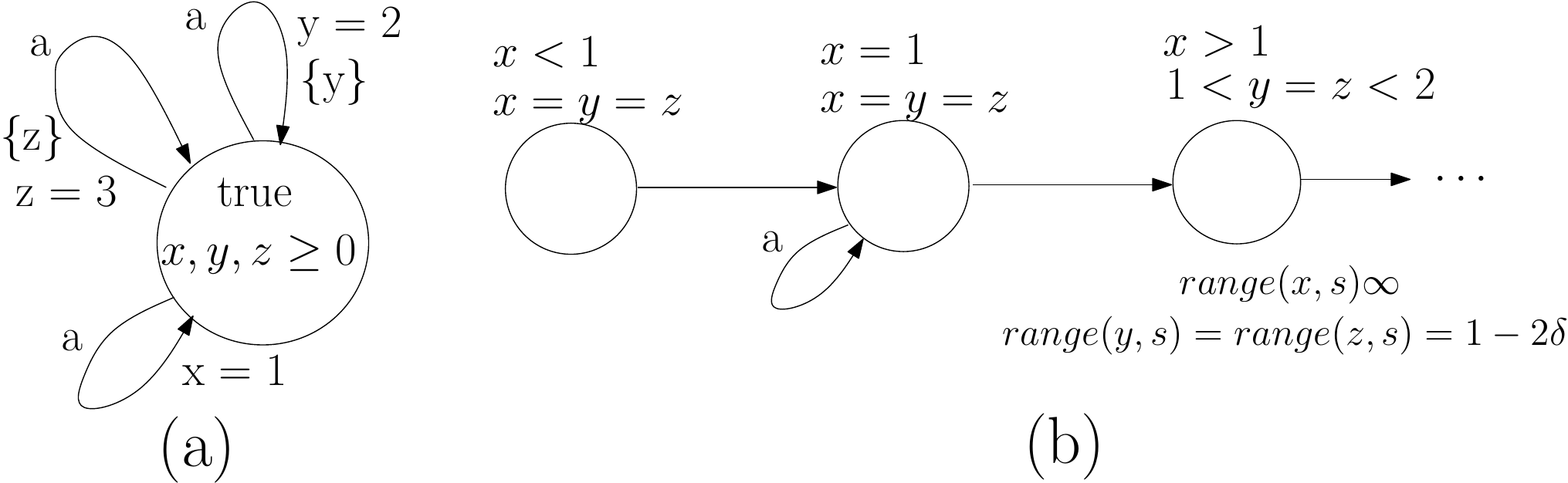}
\caption{\label{fig-range} Range of clocks in zone valuation graph node}
\end{figure}

It is possible to have different zone graphs corresponding to a timed automaton. For a timed automaton $A = (L, l_0, E, I)$ and a process $r=(l_j,v_{l_j})\in T(A)$, we are interested in a particular form of zone graph Z$_{(A,r)}$=$(S, s_r, Lep, \rightarrow)$ which satisfies the following properties:
\begin{enumerate}
\item set $S$ is finite.
\item For every node $s \in S$ the zone corresponding to the constraints $\phi_s$ is convex.
\item $v_{l_j} \models \phi_{s_r}$. Note that $v_{l_j}$ may or may not satisfy $\phi_0(C)$.
\item \label{4} For any two processes $p,q\in T(A)$, if their valuations satisfy the formula $\phi_r$ for the same node $r \in S$ then $p \sim_u q$, i.e. $p$ is time abstracted bisimilar to $q$.
\item \label{5}For two timed automata $A_1$, $A_2$ and two processes $p \in T(A_1)$ and $q \in T(A_2)$, $Z_{(A_1,p)} \sim Z_{(A_2,q)} \Leftrightarrow p \sim_u q$.
\item It should be minimal to the extent of preserving convexity of the zones and gives a canonical form.
\end{enumerate}
For any node $s \in S$, let $\mathcal{G}(s)$ represent the set of all processes reachable from $p$ with the same location as that of $s$ and whose valuations satisfy $\phi_s$. $p$ is the initial clock valuation corresponding to which the zone valuation graph is created.
The following definitions are from \cite{GNA1}.
\begin{definition}{\textbf{Span:}}
 For a given node $s \in S$ and a clock $x \in C$, $min_{x}(s)$ and $max_{x}(s)$ represent the minimum and the maximum clock valuations of a clock $x$ across all processes in node $s$. For $x \ge c$, $min_x(s) = c$, for $x > c$, $min_x(s) = c + \delta$ and $max_x(s) = \infty$. For $x \le c$, $max_x(s) = c$, for $x < c$, $max_x(s) = c - \delta$ and $min_x = 0$ in both cases. Here $\delta$ is a symbolic representation of an infinitesimally small value. We define $range(x,s)$ as $max_{x}(s)-min_{x}(s)$. The span of a node $s \in S$ is defined as $\mathcal{M}(s) = min\{range(x,s)\ |\ x\in C\}$, i.e. minimum of all clocks' ranges. We define a clock $y$ belonging to the set $\{y \:|\: range(y, s) = \mathcal{M}(s)\}$ to be a critical clock of node $s$.
\end{definition}
 For example, in a zone valuation graph with two clocks $x$ and $y$, the span of a node $s$ with $\phi_s = x > 3$ and $y < 1$ is $min(\infty,1 - \delta)=1-\delta$ whereas span for a node with $\phi_s=x > 1$ and $y = 2$ is $min(\infty,0)=0$. We say that for a node $s$ in the zone valuation graph, $range(x, s) = m- l - 2\delta$ where valuations for clock $x$ lie in the range $l < x < m$. It is to be noted that $2\delta$ is also a symbolic value. It is to be noted that for a given node $s$, $range(x,s)$ is the same for all clock variables $x \in C$ if the zone corresponding to node $s$ is not abstracted with respect to any clock variable. If the zone corresponding to $s$ is abstracted with respect to one or more clock variables then for each such variable $x \in C$, $range(x,s)=\infty$. For example in figure \ref{fig-range}, we show a timed automaton and part of its zone valuation graph. The zone corresponding to rightmost node in the part of the zone valuation graph shown inthe figure, is abstracted with respect to clock $x$ and hence $range(x, s) = \infty$, whereas $range(y,s) = range(z, s) = 1 - 2 \delta$.
\begin{definition}
Given a timed automaton $A$, let $Z_{(A, p)}$ be the zone valuation graph corresponding to process $p \in T(A)$. Let $p' \in T(A)$ be a process reachable from $p$ and $s$ be the node of $Z_{(A, p)}$ such that $p' \in \mathcal{G}(s)$. Let $x$ be a critical clock of $s$ and $v_{p'}(x)$ denote the valuation of clock $x$ for process $p'$. We define maximum admissible delay for $p'$ in $s$ as $max_{x}(s) - v_{p'}(x)$.
\end{definition}
For example, from the figure describing zone valuation graph for automaton 1 in figure \ref{fig-exTimeAbsDelayBisim} , the maximum admissible delay for the process $\langle A, x=1\rangle$ is $2-1= 1$.

The algorithm for creating zone valuation graph consists of two phases. In the first phase, forward and backward analysis of the given timed automaton produces a zone graph where zones are split based on a canonical decomposition \cite{TY1} of the constraints on the outgoing edges in the timed automaton. In the second phase, the nodes in the zone valuation graph produced after phase 1 that are strongly bisimilar to each other are merged using Paige-Tarjan algorithm \cite{PT1} to produce a canonical form of the zone valuation graph. After merging, every node in the zone valuation  graph denotes time abstracted bisimilar classes of the timed LTS of the given timed automaton that preserves convexity. Note that after phase 1, strongly bisimilar nodes corresponding to different locations of the timed automaton can also be combined. In such case, we say that the \emph{location set} of combined node is the set of locations of the nodes that are combined.

Forward analysis may cause a zone graph to become infinite \cite{BBFL1}. To ensure finiteness of the zone graph, several kinds of abstractions have been proposed in the literature \cite{DT1}\cite{BBFL1}\cite{BBLP1}. In \cite{GNA1}, \emph{location dependent maximal constants} abstraction \cite{BBFL1} is used to get a finite zone valuation graph.

The time complexity required for creation of zone valuation graph has also been derived given in \cite{GNA1}. In the worst case, the zone valuation graph created becomes same as region graph and hence the worst case complexity of creation of zone valuation graph is exponential in the number of clocks. For a given timed automaton, if $n$ be the number of locations in the timed automaton and $|S|$ and $m$ denote the number of nodes and edges respectively in the zone valuation graph produced after phase 1 of the algorithm, then the total time required in both phases for construction of the zone valuation graph is $O(n^2 (|C|^3n + |S| \times |C| +  n^2\times log\:n) + |S| \times log\:m)$.

\section{Equivalences and Preorders for Timed Systems}\label{sec-timedrels}
We discuss here several bisimulations, simulation equivalences and preorders dealing with real time for timed processes that are states or valuations of a timed automaton execution.
\begin{definition}\label{def-timedbisim}
\textbf{Timed bisimilarity}: A binary symmetric relation $\mathcal{R}_t$ over the set of states of a TLTS is a timed bisimulation relation if whenever $p_1 \mathcal{R}_t p_2$, for each action $a \in Act$ and time delay $d \in \mathbb{R}_{\ge 0}$\\[.1cm]
if $p_1 \stackrel{a}{\rightarrow} p_1'$ then there is a transition $p_2 \stackrel{a}{\rightarrow} p_2'$ such that $p_1' \mathcal{R}_t p_2'$, and \\
if $p_1 \stackrel{d}{\rightarrow} p_1'$ then there is a transition $p_2 \stackrel{d}{\rightarrow} p_2'$ such that $p_1' \mathcal{R}_t p_2'$.\\
Timed bisimilarity $\sim_t$ is the largest timed bisimulation relation.
\end{definition}
Timed automata $A_1$ and $A_2$ are \emph{timed bisimilar} if the initial states in the corresponding TLTS are timed bisimilar. Matching each time delay in one automaton with identical delays in another automaton may be too strict a requirement. Time abstracted bisimilarity is the relation obtained by a relaxation of this requirement where $p'_1 \sim_t p'_2$ is replaced uniformly by $p'_1 \sim_u p'_2$ and the second clause of definition \ref{def-timedbisim} is replaced by\\
\textit{if $p_1 \stackrel{d}{\rightarrow} p'_1$ then there is a transition  $p_2 \stackrel{d'}{\rightarrow} p'_2$, such that $p'_1 \sim_u p'_2$}.
The delay $d$ can be different from $d'$.\\[.1cm]
Timed automata $A_1$ and $A_2$ are \emph{time abstracted bisimilar} if the initial states in the corresponding TLTS are time abstracted bisimilar. 

In this work, we introduce below \emph{interval bisimulation} to bridge the gap between timed and time abstracted bisimulation and provide its game semantics later indicating how it can be decided using zone valuation graph.
\begin{definition}\label{def-intvltimedbisim}
\textbf{Interval bisimilarity}: A binary symmetric relation $\mathcal{R}_i$ over the set of states of a TLTS is an interval bisimulation relation if whenever $p_1 \mathcal{R}_i p_2$, for each action $a \in Act$ and time delays $d, \:d' \in \mathbb{R}_{\ge 0}$\\[.1cm]
if $p_1 \stackrel{a}{\rightarrow} p_1'$ then there is a transition $p_2 \stackrel{a}{\rightarrow} p_2'$ such that $p_1' \mathcal{R}_i p_2'$, and \\
if $p_1 \stackrel{d}{\rightarrow} p_1'$ then there is a transition $p_2 \stackrel{d'}{\rightarrow} p_2'$ such that $p_1' \mathcal{R}_i p_2'$ and $d' = d$ if $frac(d) = 0$ and $d' \in (\:\lfloor d \rfloor, \lceil d \rceil\:)$ otherwise. Here frac(d) denotes the fractional part of delay $d$.\\
Interval bisimilarity $\sim_i$ is the largest interval bisimulation relation.
\end{definition}
\begin{definition}\label{def-delaybisim}
\textbf{Time Abstracted Delay Bisimilarity}: A binary symmetric relation $\mathcal{R}_y$ over the set of states of a TLTS is a time abstracted delay bisimulation relation if whenever $p_1 \mathcal{R}_y p_2$, for each action $a \in Act$ and time delays $d, \:d' \in \mathbb{R}_{\ge 0}$\\[.1cm]
if $p_1 \stackrel{a}{\rightarrow} p_1'$ then there is a transition $p_2 \stackrel{d}{\rightarrow} \stackrel{a}{\rightarrow} p_2'$ such that $p_1' \mathcal{R}_y p_2'$, and \\
if $p_1 \stackrel{d}{\rightarrow} p_1'$ then there is a transition $p_2 \stackrel{d'}{\rightarrow} p_2'$ such that $p_1' \mathcal{R}_y p_2'$.\\
Time abstracted delay bisimilarity $\sim_y$ is the largest time abstracted delay bisimulation relation.
\end{definition}
\begin{definition}
A time abstracted observational bisimulation relation, $\mathcal{R}_o$ can be defined by replacing $\mathcal{R}_y$ uniformly with $\mathcal{R}_o$ in definition \ref{def-delaybisim} and the first clause in definition \ref{def-delaybisim} being replaced by the following: \\
\emph{if $p_1 \stackrel{a}{\rightarrow} p_1'$ then there is a transition $p_2 \stackrel{d_1}{\rightarrow} \stackrel{a}{\rightarrow} \stackrel{d_2}{\rightarrow} p_2'$ such that $p_1' \mathcal{R}_o p_2'$}.
Time abstracted observational bisimilarity, denoted by $\sim_o$, is the largest time abstracted observational bisimulation relation.
\end{definition}
\begin{definition}\label{def-timedsim}
\textbf{Timed Simulation}: A timed process $p_2$ is said to time simulate process $p_1$ if there exists a relation $\mathcal{R}_1$ such that for each action $a \in Act$ and time delay $d \in \mathbb{R}_{\ge 0}$\\[.1cm]
if $p_1 \stackrel{a}{\rightarrow} p_1'$ then there is a transition $p_2 \stackrel{a}{\rightarrow} p_2'$ such that $p_1' \mathcal{R}_1 p_2'$, and \\
if $p_1 \stackrel{d}{\rightarrow} p_1'$ then there is a transition $p_2 \stackrel{d}{\rightarrow} p_2'$ such that $p_1' \mathcal{R}_1 p_2'$.\\
$p_1$ and $p_2$ are said to be timed simulation equivalent if $p_1$ time simulates $p_2$ and $p_2$ time simulates $p_1$.
\end{definition}
Thus corresponding to each of the bisimulation relation defined above, we can define a simulation equivalence.

The following definition of timed performance prebisimulation is from \cite{GNA1}.
\begin{definition}\label{def-timedprebisim}
\textbf{Timed performance prebisimilarity}: A binary relation $\mathcal{B}$ over the set of states of a TLTS is a \emph{timed performance prebisimulation relation} if whenever $p_1 \mathcal{B} p_2$, for each action $a \in Act$ and time delay $d \in \mathbb{R}_{\ge 0}$\\[.1cm]
if $p_1 \stackrel{a}{\rightarrow} p_1'$ then there is a transition $p_2 \stackrel{a}{\rightarrow} p_2'$ such that $p_1' \mathcal{B} p_2'$, and \\
if $p_2 \stackrel{a}{\rightarrow} p_2'$ then there is a transition $p_1 \stackrel{a}{\rightarrow} p_1'$ such that $p_1' \mathcal{B} p_2'$, and \\
if $p_1 \stackrel{d}{\rightarrow} p_1'$ then there is a transition $p_2 \stackrel{d'}{\rightarrow} p_2'$ for $d \le d'$ such that $p_1' \mathcal{B} p_2'$ ,and \\
if $p_2 \stackrel{d}{\rightarrow} p_2'$ then there is a transition $p_1 \stackrel{d'}{\rightarrow} p_1'$ for $d \ge d'$ such that $p_1' \mathcal{B} p_2'$.\\
Timed performance prebisimilarity $\precsim$ is the largest timed performance prebisimulation relation.
\end{definition}

\subsection{Comparison Among these Relations}
It is easy to see from the definitions that strong timed bisimulation implies strong time-abstracted bisimulation whereas the converse is not true. Interval bisimulation lies in between timed bisimulation and time abstracted bisimulation and from the definitions, $\sim_u \:\subseteq\: \sim_y \:\subseteq \:\sim_o$. Also existence of a bisimulation relation between two processes implies the existence of the corresponding simulation equivalence. It is also easy to see that timed performance prebisimulation lies in between timed bisimulation and time abstracted bisimulation. Though not immediately evident, we will subsequently prove that timed performance prebisimulation is weaker than interval bisimulation. In figure \ref{fig-timedSpectrum}, an arrow from one relation to the other denotes that the relation from which the arrow originates is stronger than the one to which it points. Hence we have $\sim_t \:\Rightarrow \:\sim_i \:\Rightarrow \:\precsim \:\Rightarrow \:\sim_u \:\Rightarrow \:\sim_y \:\Rightarrow \:\sim_o$ and it is easy to see that similar implication relations also exist among the corresponding simulation equivalences. Thus we obtain figure $\ref{fig-timedSpectrum}$ where $\mathcal{R}_1 \longrightarrow \mathcal{R}_2$ denotes that $\mathcal{R}_1$ is a strict subset of $\mathcal{R}_2$.

\section{Game Characterization}\label{sec-timedgames}
In \cite{CD1}, a hierarchy of games has been proposed that allows systematic comparison of process equivalences for discrete processes. The process hierarchy of Van Glabbeek can be embedded in the game hierarchy defined in \cite{CD1}. In this work we provide a similar game hierarchy so as to correspond to process equivalences and preorders that involve real time. Similar to the games in \cite{CD1}, our games are also Ehrenfeucht-Fra\"{\i}ss\'{e} games where player I is known as the \emph{attacker} and player II is called the \emph{defender}. The game is played on a finite graph. In our case this finite graph is either the zone valuation graph or one of its variants as described later in detail. Corresponding to the two timed processes for which we want to check if they are related through one of the relations described in section \ref{sec-timedrels}, two graphs are first created on which the game is to be played. As in every EF game, the attacker chooses a graph and makes its move. The defender tries to replicate the move on the other graph. If the defender can always replicate the move the attacker makes, then it wins implying that the two processes are related through the relation that corresponds to the game. If at any point in time, the defender cannot replicate the move of the attacker, then it loses which implies that the two processes are not related through the corresponding relation. In a bisimulation game before any round, the attacker can also choose the graph on which it will make its move. The defender has to choose the other graph. If the attacker changes the graph between two consecutive rounds, it is known as an \emph{alternation}. Alternations are not allowed in games corresponding to simulation equivalences. A game can be played infinitely or for a finite number of rounds. The moves made by the attacker or the defender can also differ from one game to another. In the EF games described in this section, the moves denote an action or a sequence of actions belonging to the set $Act \cup \{\varepsilon\}$. Certain extra conditions can also be part of the game depending on the relation to which the game corresponds to. For example, in timed bisimulation game, after every move the defender needs to ensure that the span of its current node is exactly same as the span of the node in which the attacker resides. Ensuring the equality of the span is an extra condition.

\subsection{Game Template}
A timed game proposed in this work can be described using the grammar $\mathcal{L} ::= n-\Gamma^{G, \alpha, \beta}_{k} \;, \; \mathcal{L}_1 \vee \mathcal{L}_2$. Each game is characterized by the following parameters as described below:
\begin{itemize}
\item $n$ : number of alternations. If not mentioned, it denotes no restriction on the number of alternations in the subgame.
\item $k$ : number of rounds; a subgame can have even infinite number of rounds.
\item $G$ : underlying graph on which the game is played. It can be of the following types: $Z$ denotes zone valuation graph, $Z_1$ denotes the graph obtained after phase 1 of zone valuation graph construction. This can be used for games of time abstracted relations. $Z_{sim}$ denotes the graph that is obtained by combining the nodes of $Z_1$ that are simulation equivalent.
\item $\alpha$ : a vector of two elements: the first element denotes the move chosen by attacker whereas the second element denotes the move chosen by defender.
\item $\beta$ : extra condition in the game and may be of the following types:
\begin{itemize}
\item $=$ : This condition denotes that span has to be matched. We also use $(s_1 = s_2)$ to denote that the spans of nodes $s_1$ and $s_2$ should be the same.
\item $\lfloor=\rfloor$ : This condition denotes that the integer portion of the span has to be matched and if the decimal part of one span is 0, then so should be for the other. We also sometimes use $(s_1 \lfloor = \rfloor s_2)$ where $s_1$ and $s_2$ are nodes of the two zone valuation graphs.
\item $G_1, \le$ : Let the two graphs for the two timed processes be denoted by $G_1$ and $G_2$. This extra condition denotes that the span of any node in $G_1$ should be less than or equal to the span of the corresponding bisimilar node in $G_2$.
\end{itemize}
$\beta$ if not specified denotes that there is no extra condition.
\end{itemize}

\subsection{Time Abstracted Bisimulation Game}
This is the EF bisimulation game played on the zone valuation graphs of two timed processes. There is no restriction on the number of rounds and the number of alternations.
\begin{lemma}
The game $\Gamma^{Z, \langle a, a \rangle}_{\infty}$, where $a \in Lep$ characterizes time abstracted bisimulation.
\end{lemma}
\proof This is a strong bisimulation game played on two zone valuation graphs. By construction of zone valuation graph, two processes are time abstracted bisimilar if their corresponding zone valuation graphs are strongly bisimilar. Hence the proof.
\qed
Note that for any kind of time abstracted relation, the zone graph obtained after phase 1 of the zone valuation graph creation algorithm can be used. The intuition behind this is that in phase 2, the zones that are behaviorally similar (bisimilar or simulation equivalent) are combined in this phase. Only the span of the combined zone changes which is required for matching the time. Thus phase 2 is important for timed relations only.
\begin{example}
Figure \ref{fig-exTimeAbsBisim} shows two timed automata and their corresponding zone valuation graphs for timed processes $\langle A, x = 0 \rangle$ and $\langle A',x =  0\rangle$. The defender has a universal winning strategy for the game $\Gamma^{Z, \langle a, a \rangle}_{\infty}$ and hence the two processes are time abstracted bisimilar.
\begin{figure}
\centering
\includegraphics[width=0.7\textwidth]{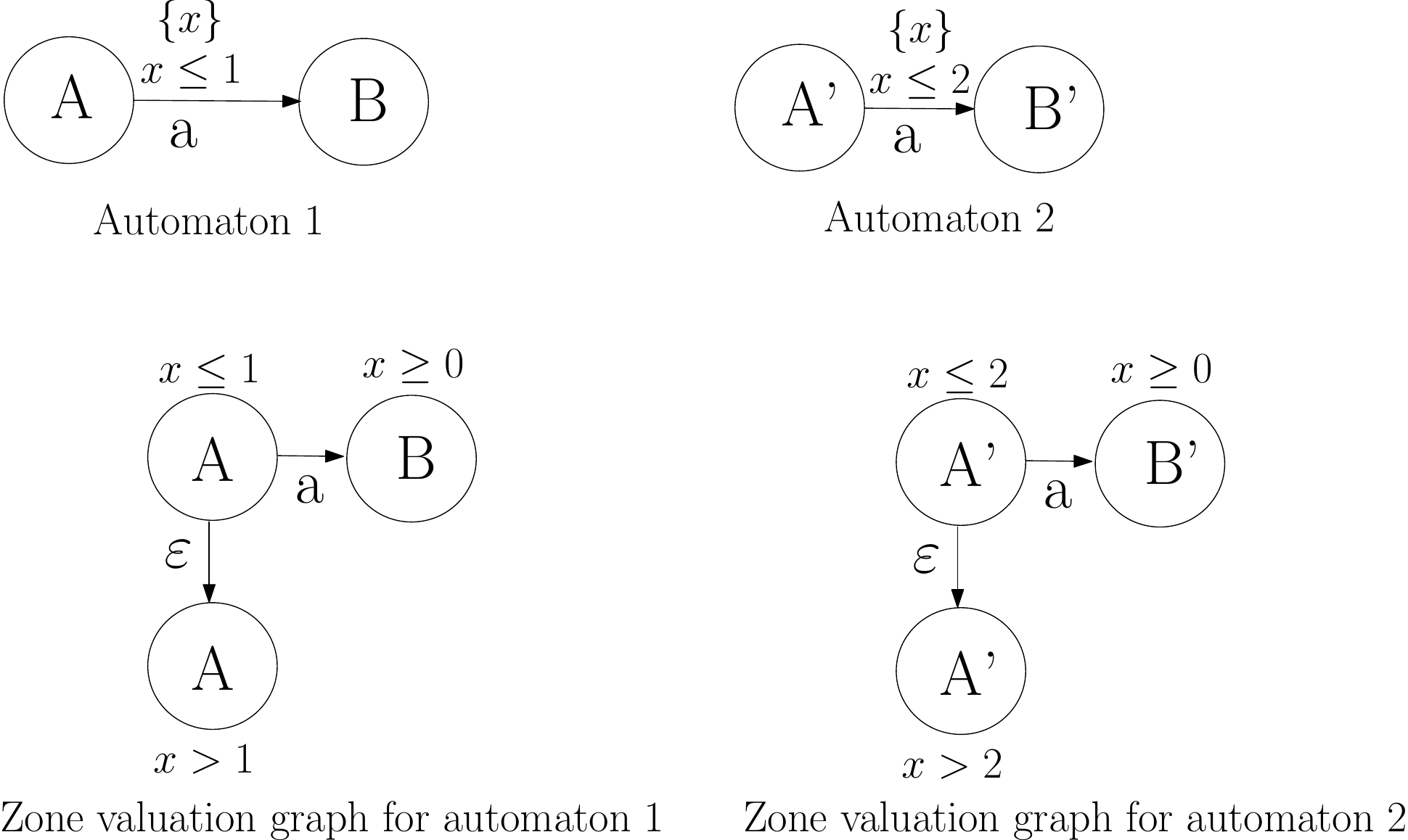}
\caption{\label{fig-exTimeAbsBisim} Example of time abstracted bisimulation game}
\end{figure}	
\end{example}

\subsection{Timed Bisimulation Game}
This game is same as the game for time abstracted bisimulation but has an extra condition which specifies that the spans of every pair of bisimilar nodes from the two zone valuation graphs should be equal.
\begin{lemma}
The game $\Gamma^{Z, \langle a, a,\rangle, =}_{\infty}$, where $a \in Lep$ characterizes timed bisimulation.
\end{lemma}
\proof In this game if the defender has a universal winning strategy then it implies that the two zone valuation graphs are strongly bisimilar and every pair of bisimilar nodes in the two zone valuation graphs have equal span. This implies that the two timeed processes are timed bisimilar. The detailed proof is given in \cite{GNA2}.
\qed
\begin{example}
In this example, we consider two timed automata as given in \cite{LAKJ1}. Figure \ref{fig-exTimedBisim} shows the two timed automata and their corresponding zone valuation graphs for timed processes $\langle A, x = 0 \rangle$ and $\langle A', x = 0 \rangle$. The defender has a universal winning strategy for the game $\Gamma^{Z, \langle a, a \rangle, =}_{\infty}$ and hence the two processes are timed bisimilar. In  the figure, the spans of the nodes are indicated within parentheses.
\begin{figure}
\centering
\includegraphics[width=0.6\textwidth]{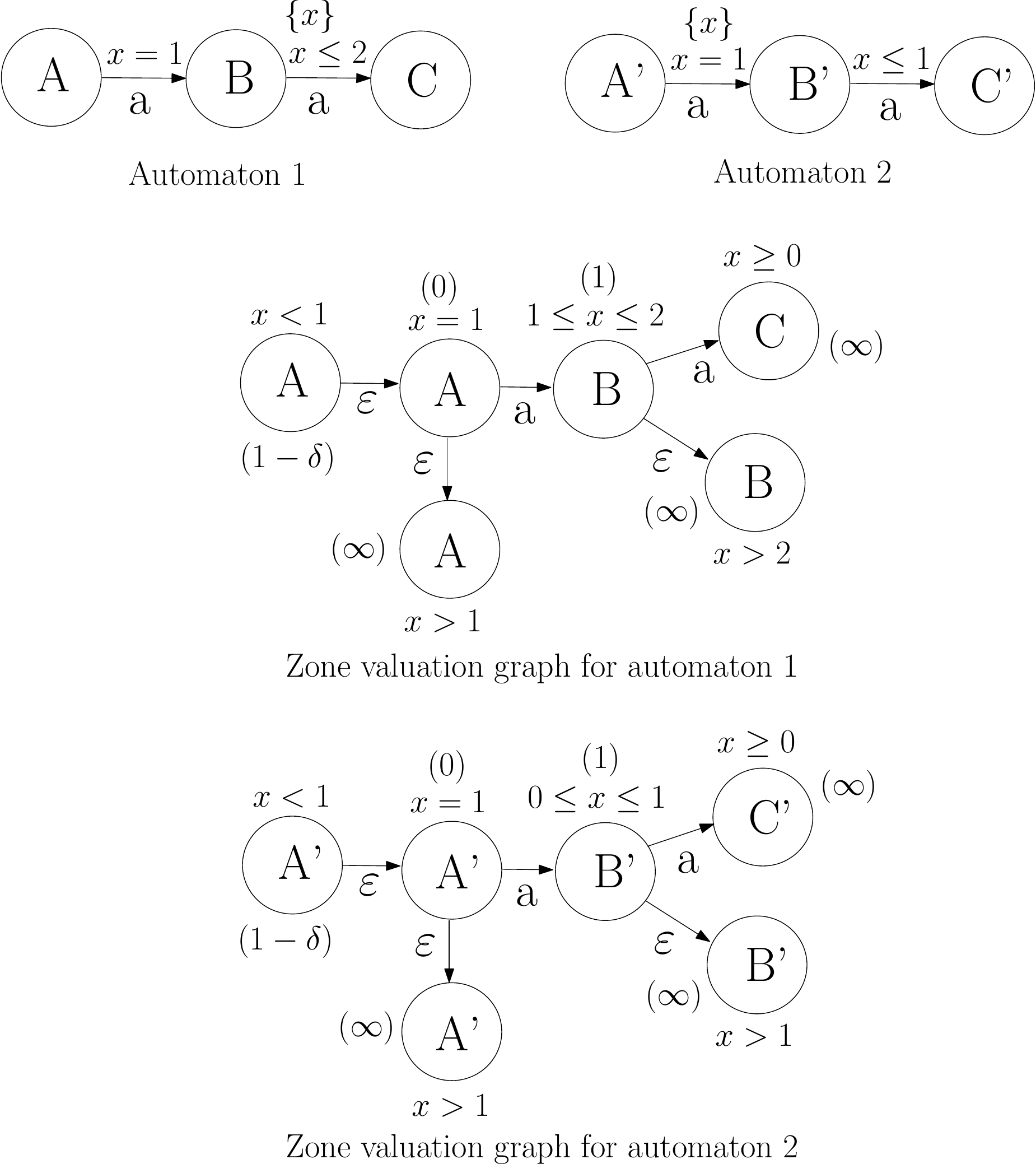}
\caption{\label{fig-exTimedBisim} Example of timed bisimulation game}
\end{figure}	
\end{example}

\subsection{Interval Bisimulation Game}
This game is same as the game for timed bisimulation with the following difference. Let $s_p$ and $s_q$ be the initial nodes of the zone valuation graphs corresponding to processes $p$ and $q$. It is not required that the spans of $s_p$ and $s_q$ have to be equal but the integer parts of the spans should be the same and if the fractional part of one span is 0, so should be for the other node. Thus the game characterization for interval bisimulation is $\Gamma^{Z, \langle a, a \rangle, (s_p \lfloor = \rfloor s_q, s_1 = s_2)}_\infty$, where $(s_1, s_2) \neq (s_p, s_q)$ and $a \in Lep$.

\begin{theorem} \label{thm-interval}
A universal winning strategy for the defender in the game $\Gamma_{\infty}^{Z, \langle a, a\rangle, (s_p \lfloor = \rfloor s_q, s_1 = s_2)}$, where $(s_1, s_2) \neq (s_p, s_q)$ and $a \in Lep$ denotes that the two timed processes $p$ and $q$ are interval bisimilar. Here $s_p$ and $s_q$ are the initial nodes of the two zone valuation graphs.
\end{theorem}
\proof
For the initial nodes $s_p$ and $s_q$, $\lfloor \mathcal{M}(s_p)\rfloor  = \lfloor\mathcal{M}(s_q) \rfloor$, i.e. the integer portions of the spans match and $frac(\mathcal{M}(s_p)) = 0 \Leftrightarrow frac(\mathcal{M}(s_q)) = 0$ and for the rest of the bisimilar nodes from the two zone valuation graphs, their spans are equal $\Rightarrow \; \sim_i$. This implication is easy to see.\\[.2cm]
$\sim_i \; \Rightarrow$ for initial nodes $s_p$ and $s_q$, $frac(\mathcal{M}(s_p)) = 0 \Leftrightarrow frac(\mathcal{M}(s_q)) = 0$ and $\lfloor \mathcal{M}(s_p)\rfloor  = \lfloor\mathcal{M}(s_q) \rfloor$, i.e. the integer portions of the spans match and for the rest of the bisimilar nodes from the two zone valuation graphs, their spans have to be equal.

We prove this below. Considering the initial nodes, there can be two cases:
\begin{enumerate}
\item $frac(\mathcal{M}(s_{p}))$ = 0. By the definition of interval bisimulation $\mathcal{M}(s_{q}) = \mathcal{M}(s_{p})$.
\item when $frac(\mathcal{M}(s_{p})) \neq 0$. From the definition of interval bisimulation, this also requires that $frac(\mathcal{M}(s_{q})) \neq 0$.

Also it is straightforward to see that for $p$ and $q$ to be interval bisimilar, $\lfloor frac(\mathcal{M}
(s_{p})) \rfloor = \lfloor frac(\mathcal{M}(s_{q})) \rfloor$, i.e. their integer parts are the same. We 
can prove this by contradiction. Suppose without loss of generality, the integer parts of the spans of $s_p$ and $s_q$ are respectively $t$ and $t + l$, where $t$ and $l$ are positive integers. Thus $p$ can make a delay $d = t+1$ to become $p'$ whereas $q$ cannot make a delay $t+1$ such that $q \stackrel{d=t+1}{\longrightarrow}{q'}$ and $p' \sim_i q'$ since such a $p' \not \in \mathcal{G}(s_p)$ whereas $q' \in \mathcal{G}(s_q)$.
\end{enumerate}
For the bisimilar nodes apart from the pair of initial nodes in the two zone valuation graphs, the spans have to be exactly same. The span of a node can be of the forms $t$, $t - \delta$ or $t - 2 \delta$, where $\delta$ symbolizes an infinitesimally small number.

Exactly with the same argument as above, we can show that two processes $p$ and $q$ cannot be interval bisimilar if any two bisimilar nodes in their corresponding zone graphs have spans $t$ and $t+l$, where $t$ and $l$ are positive integers.

Now we consider the case where the spans of two bisimilar nodes are $t$ and $t - \delta$. Let $p \stackrel{tr} {\longrightarrow} p'$, where $tr \in Lep^+$ and $p' \in \mathcal{G}(s_{p'})$ and $\mathcal{M}(s_{p'}) = t$. Similarly, let us suppose $q \stackrel{tr} {\longrightarrow} q'$ and $q' \in \mathcal{G}(s_{q'})$ and $\mathcal{M}(s_{q'}) = t - \delta$ and $s_{p'}$ and $s_{q'}$ form the pair of bisimilar nodes. We prove that in such a case $p$ and $q$ are not interval bisimilar.

Let in the paths from $s_p$ to $s_{p'}$ and from $s_q$ to $s_{q'}$, $s_{p_1}$ and $s_{q_1}$ be the first pair of nodes that are strongly bisimilar to each other such that the spans of $s_{p_1}$ and $s_{q_1}$ be $m$ and $m - \delta$ respectively. It is possible that $s_{p_1}$ is same as $s_{p'}$ and $s_{q_1}$ is same as $s_{q'}$. There can be two cases which can cause the span of $s_{q_1}$ to be $m - \delta$.
\begin{enumerate}
\item Lower limit of value of the critical clock $y$ is $j + \delta$ and the upper limit being $j + m$ where $j$ is an integer.
\item Lower limit of value of the critical clock $y$ is the integer $j$ and the upper limit being $j + m - \delta$.
\end{enumerate}
We start with the first case. Let $p_1 \in \mathcal{G}(s_{p_1})$ be the process such that $min_y(s_{p_1}) = v_{p_1}(y) = j$. Now we consider the transitions from $p$ to $p_1$ by delays of 1 time unit interspersed with visible action transitions. Process $q$ being interval bisimilar to $p$, performs the same actions. The delays of 1 time unit by the $p$-derivatives are exactly matched by the $q$-derivatives. 

However, process $q$ by executing the same trace as executed by $p$ to evolve into $p_1$ will not lead into a process belonging to $\mathcal{G}(s_{q_1})$ since the valuation of every clock of the q derivative by executing the trace will be an integer and will not be of the form $j + \delta$. Thus $p$ and $q$ are not interval bisimilar if the lower limit of the valuation of their critical clocks are both not integers.

We can also prove similarly for the second case too that $p$ and $q$ will not be interval bisimilar.

Now let us consider the case where the spans of two bisimilar nodes are of the form $m$ and $m - 2 \delta$. Similar to the proof of the case where the spans are $m$ and $m - \delta$, it can be proved that processes $p$ and $q$ are not interval bisimilar. The proof for the case where the spans are of the form $m-\delta$ and $m  - 2\delta$ is also very similar.
\qed
\begin{corollary}
$p \sim_i q \Rightarrow p \precsim q$, where $p$ and $q$ are two timed processes.
\end{corollary}
\proof Suppose $p$ and $q$ are interval bisimilar and let their zone valuation graphs be $Z_{A_1, p}$ and $Z_{A_2, q}$ respectively with initial nodes $s_p$ and $s_q$. Without loss of generality, say $\mathcal{M}(s_p) \ge \mathcal{M}(s_q)$. Let $\mathcal{B}$ be a strong bisimulation relation such that for $(s_p, s_q) \in \mathcal{B}$, $\mathcal{M}(s_p) \ge \mathcal{M}(s_q)$ and for the rest of the pairs of bisimilar nodes in $\mathcal{B}$, their spans are equal. This implies that $\mathcal{B}$ is a timed performance prebisimulation relation.
\qed
\begin{example}
Figure \ref{fig-exTimeIntervalBisim} (a)  and (b) show two timed automata processes $\langle A, 2.4\rangle$ and $\langle A', 0.8\rangle$ and their corresponding zone valuation graphs in (c) and (d) respecively. The defender has a universal winning strategy for the game $\Gamma^{Z, \langle a, a \rangle, (s_p \lfloor = \rfloor s_q, s_1 = s_2)}_\infty$, where $(s_1, s_2) \neq (s_p, s_q)$ and $a \in Lep$ and hence the two processes are interval bisimilar Here $s_p$ and $s_q$ are the initial nodes of the zone valuation graphs corresponding to processes $\langle A, 2.4\rangle$ and $\langle A', 0.8\rangle$. Note that the two timed automata states $\langle A, 2.4\rangle$ and $\langle A', 0.8\rangle$ 
\begin{figure}
\centering
\includegraphics[width=0.7\textwidth]{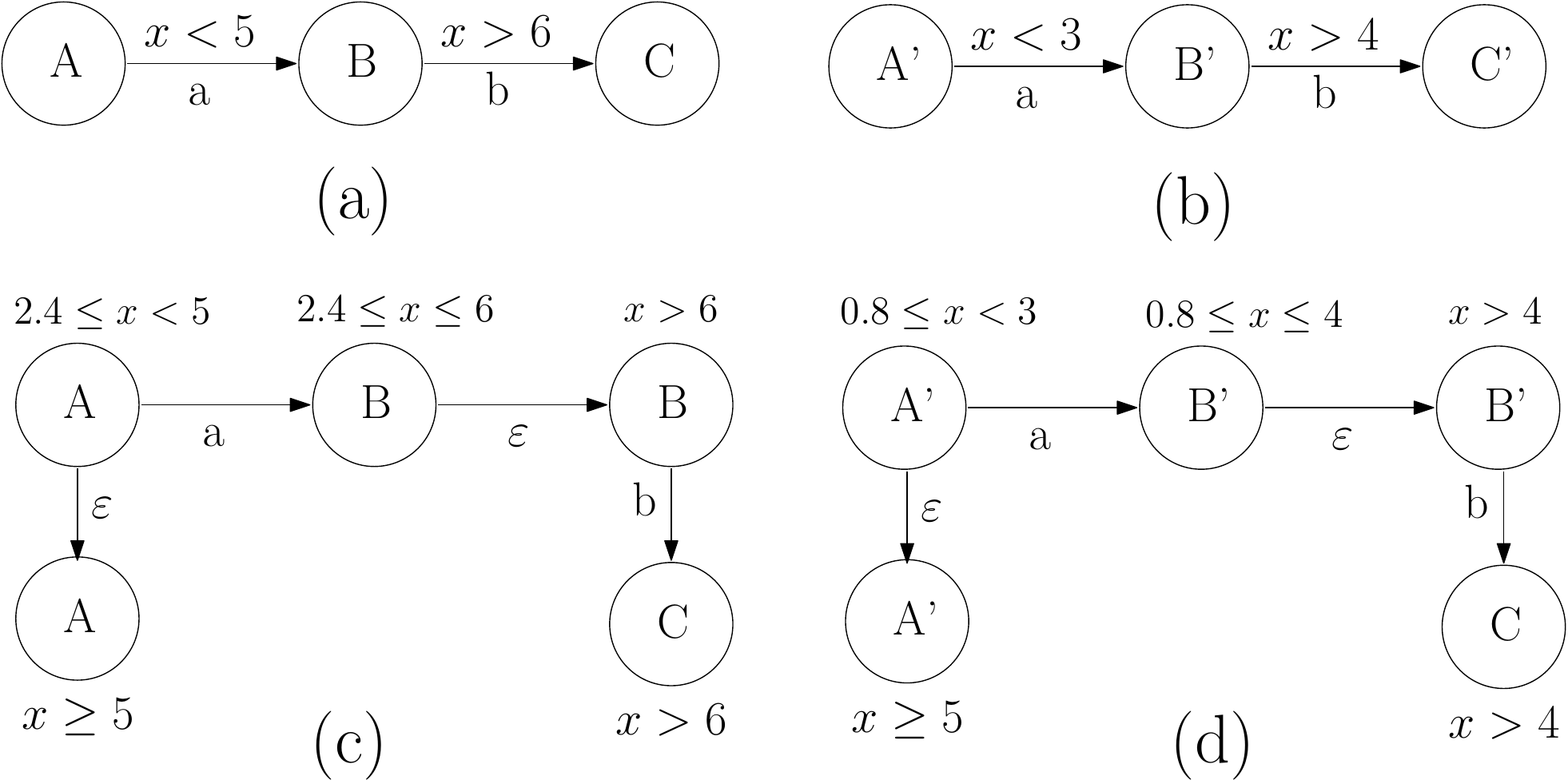}
\caption{\label{fig-exTimeIntervalBisim} Figures (c) and (d) are zone valuation graphs for states $\langle A, 2.4\rangle$ and $\langle A', 0.8\rangle$ respectively}
\end{figure}	
\end{example}

\subsection{Time Abstracted Delay Bisimualtion Game}
\begin{lemma}
The game $\Gamma^{Z, \langle a, \:\varepsilon \rightarrow a \rangle}_{\infty}$, where $a \in Lep$ characterizes time abstracted delay bisimualtion.
\end{lemma}
\proof: Since $\varepsilon$ in the graph represents a process delay, it is immediate from the definition of time abstracted delay bisimualtion.
\qed
\begin{example}
Figure \ref{fig-exTimeAbsDelayBisim} shows two timed automata and their corresponding zone valuation graphs for timed processes $\langle A, 0\rangle$ and $\langle A', 0\rangle$. $\langle A', 0\rangle$ can perform an $a$ action whereas $\langle A, 0\rangle$ can perform $a$ after performing an $\varepsilon$. The defender has a universal winning strategy for the game $\Gamma^{Z, \langle a, \:\varepsilon \rightarrow a \rangle}_{\infty}$ and hence the two processes are time abstracted delay bisimilar.
\begin{figure}
\centering
\includegraphics[width=0.7\textwidth]{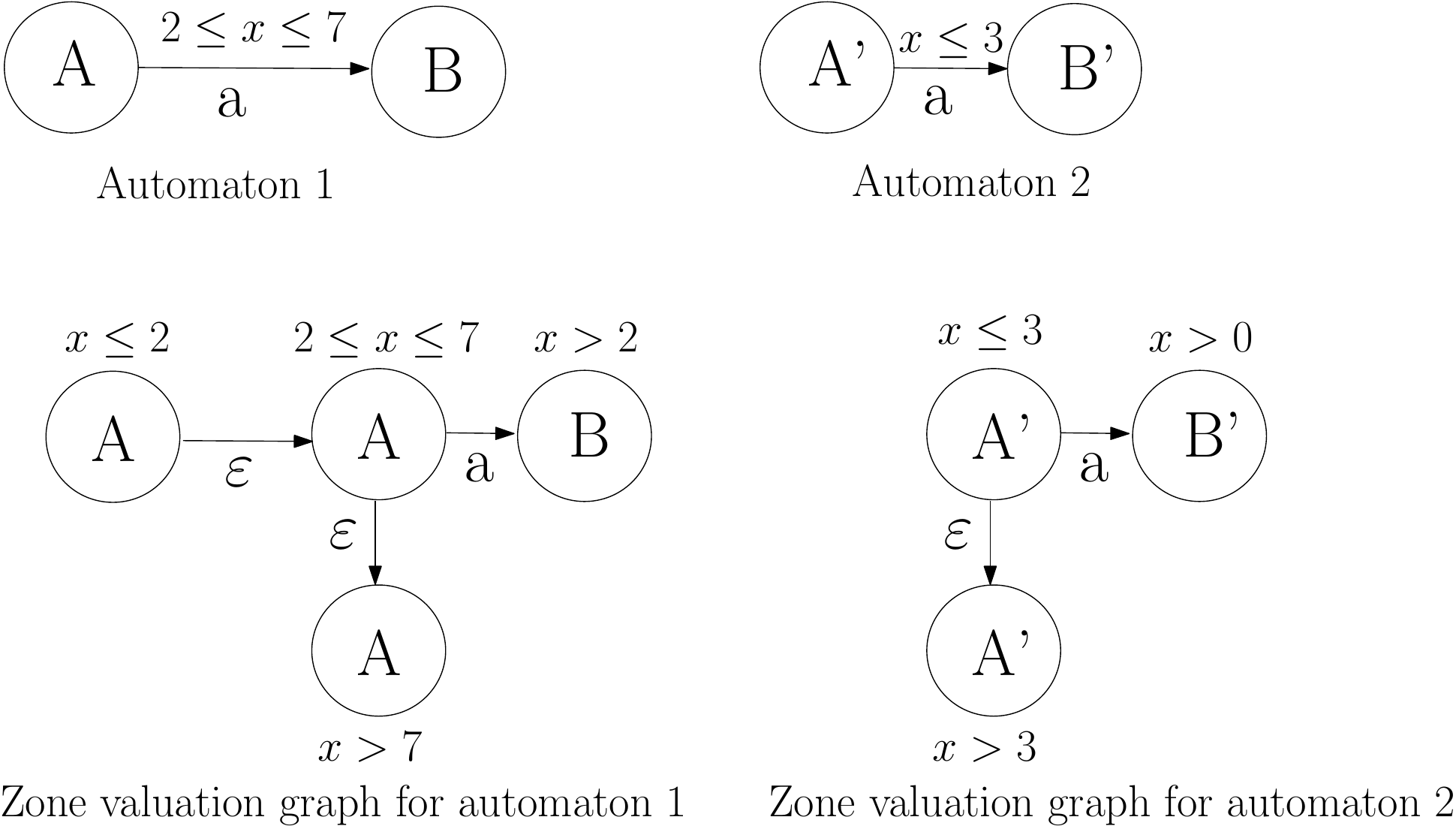}
\caption{\label{fig-exTimeAbsDelayBisim} Example of time abstracted delay bisimulation game}
\end{figure}	
\end{example}

\subsection{Time Abstracted Observational Bisimulation Game}
\begin{lemma}
The game $\Gamma^{Z, \langle a, \:\varepsilon \rightarrow a \rightarrow \varepsilon \rangle}_{\infty}$ where $a \in Lep$ characterizes time abstracted observational bisimualtion.
\end{lemma}
\proof Immediate from the definition of time abstracted observational bisimulation game.
\qed
From the definition, this game can be defined as $\Gamma^{Z, \langle a, \:\varepsilon \rightarrow a \rightarrow \varepsilon \rangle}_{\infty}$ where $a \in Lep$.
\begin{example}
In figure \ref{fig-timeAbsObs}, two timed automata from \cite{TY1} are shown that are time abstracted observation bisimilar but not time abstracted delay bisimilar. Figure \ref{fig-timeAbsObsZone} shows the corresponding zone valuation graphs and we can see that the defender has a universal winning strategy for the game $\Gamma^{Z, \langle a, \:\varepsilon \rightarrow a \rightarrow \varepsilon \rangle}_{\infty}$.
\begin{figure}
\centering
\includegraphics[width=0.6\textwidth]{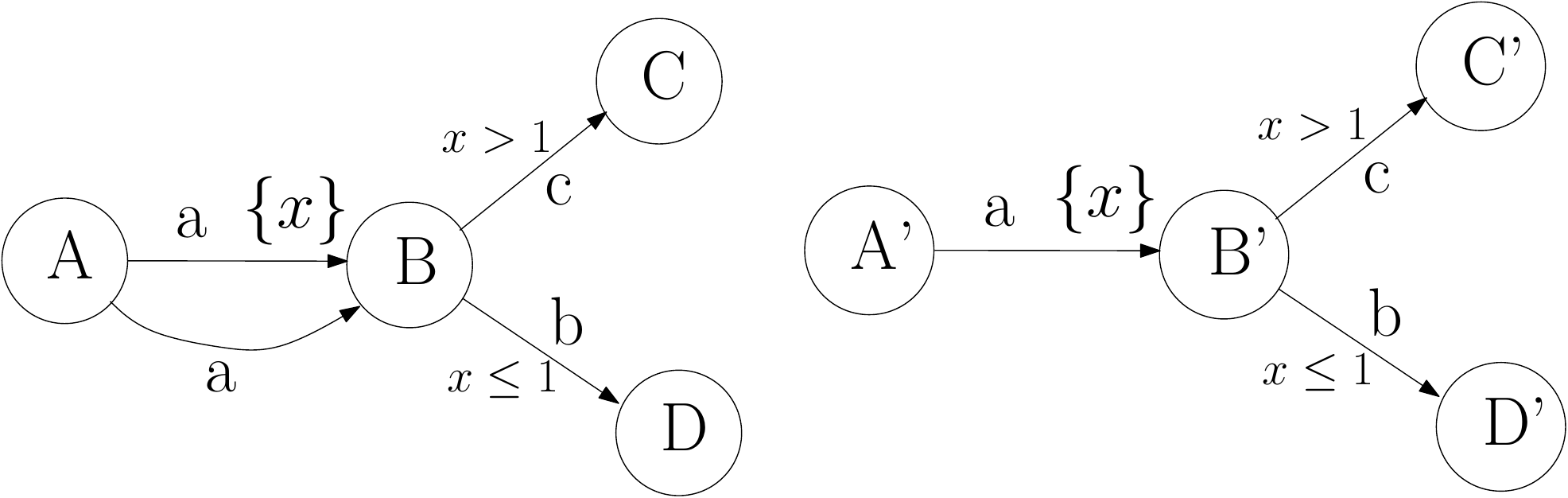}
\caption{\label{fig-timeAbsObs} A and A' are time abstracted observation bisimilar but not time abstracted delay bisimilar}
\end{figure}	
\begin{figure}
\centering
\includegraphics[width=0.7\textwidth]{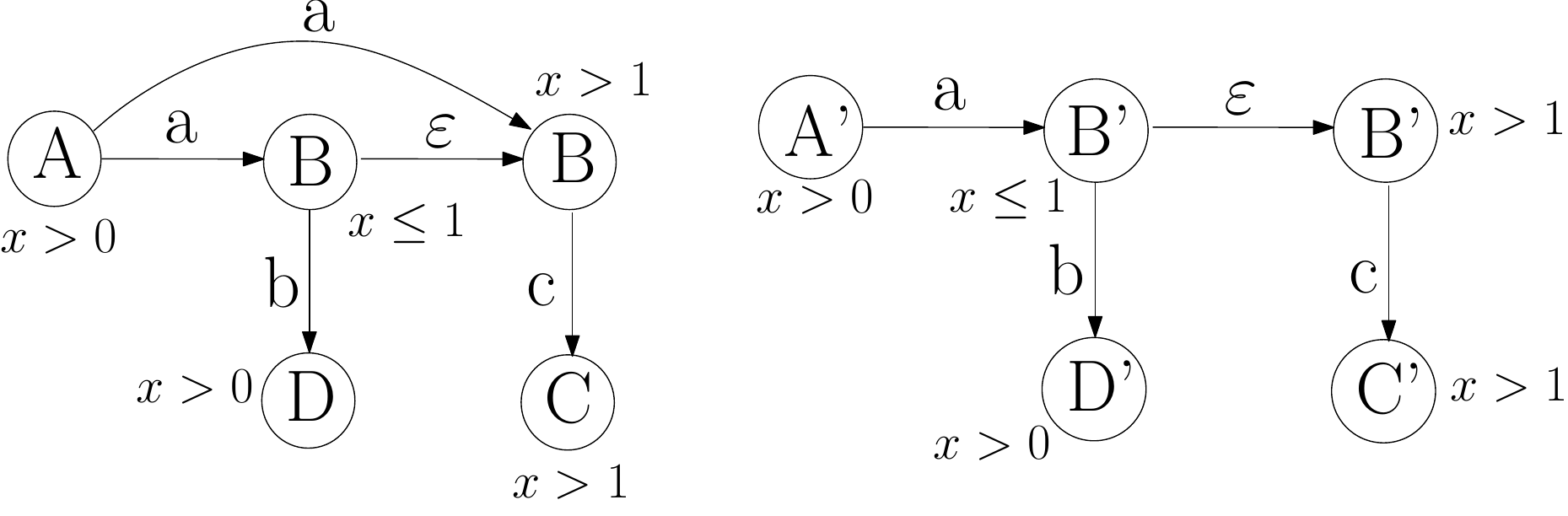}
\caption{\label{fig-timeAbsObsZone} Time abstracted observation bisimulation game for automata shown in figure \ref{fig-timeAbsObs}}
\end{figure}	
\end{example}

\subsection{Time Abstracted Simulation Equivalence Game}
This game is similar to that of time abstracted bisimulation but dos not involve any alternation.
\begin{lemma}
The game is $0-\Gamma^{Z, \langle a, a \rangle}_{\infty}$ where $a \in Lep$ characterizes time abstracted simulation equivalence.
\end{lemma}
\proof Time abstracted simulation equivalence game can be considered to be a discrete simulation equivalence game which is a discrete bisimulation game without any alternation. Hence the proof.
\qed
Note that this game can also be played on the following zone graphs.
\begin{enumerate}
\item Like other time abstracted games, the zone graph $Z_1$ obtained after phase 1 of zone valuation graph generation.
\item A phase 2 can be executed, but in stead of combining the nodes that are strongly bisimilar to each other, the nodes that are simulation equivalent to each other are combined to get a canonical form of the zone valuation graph, where the nodes denote simulation equivalent classes of the timed automata valuations.
\end{enumerate}
On similar lines, we can also define the games for time abstracted delay bisimulation equivalence and time abstracted observational bisimulation equivalence as $0-\Gamma_{\infty}^{Z, \langle a,\: \varepsilon \rightarrow a\rangle}$ and $0-\Gamma_{\infty}^{Z, \langle a,\: \varepsilon \rightarrow a \rightarrow \varepsilon \rangle}$ respectively, where $a \in Lep$.

\subsection{Timed Simulation Equivalence Game}
Designing this game is tricky when the equivalence includes real time. In the untimed domain as in \cite{CD1}, a simulation equivalence game can be obtained from the bisimulation game by restricting the number of alternations to 0. In the timed version though, this is not the case. Thus the game $0-\Gamma^{Z, \langle a, a \rangle, =}_{\infty}$ where $a \in Lep$ does \emph{not} characterize timed simulation equivalence. This can be shown with the following example:
\begin{example}
Figure \ref{fig-exTimedSimEquiv} shows two timed automata and their corresponding zone valuation graphs for timed processes $\langle A, x= 0\rangle$ and $\langle A', x= 0\rangle$. In the first zone valuation graph, corresponding to location $A$, the nodes that are created are named $A_1$, $A_2$ and $A_3$ for convenience. The two processes are timed simulation equivalent though the defender does not have a universal winning strategy in the game $0-\Gamma^{Z, \langle a, a \rangle, =}_{\infty}$ as the spans of $A_1$ and $A'$ do not match. Note that here $A_1$ and $A_2$ are not strongly bisimilar and hence cannot be merged while creating the canonical form of the zone valuation graph through phase 2. 
\begin{figure}
\centering
\includegraphics[width=0.7\textwidth]{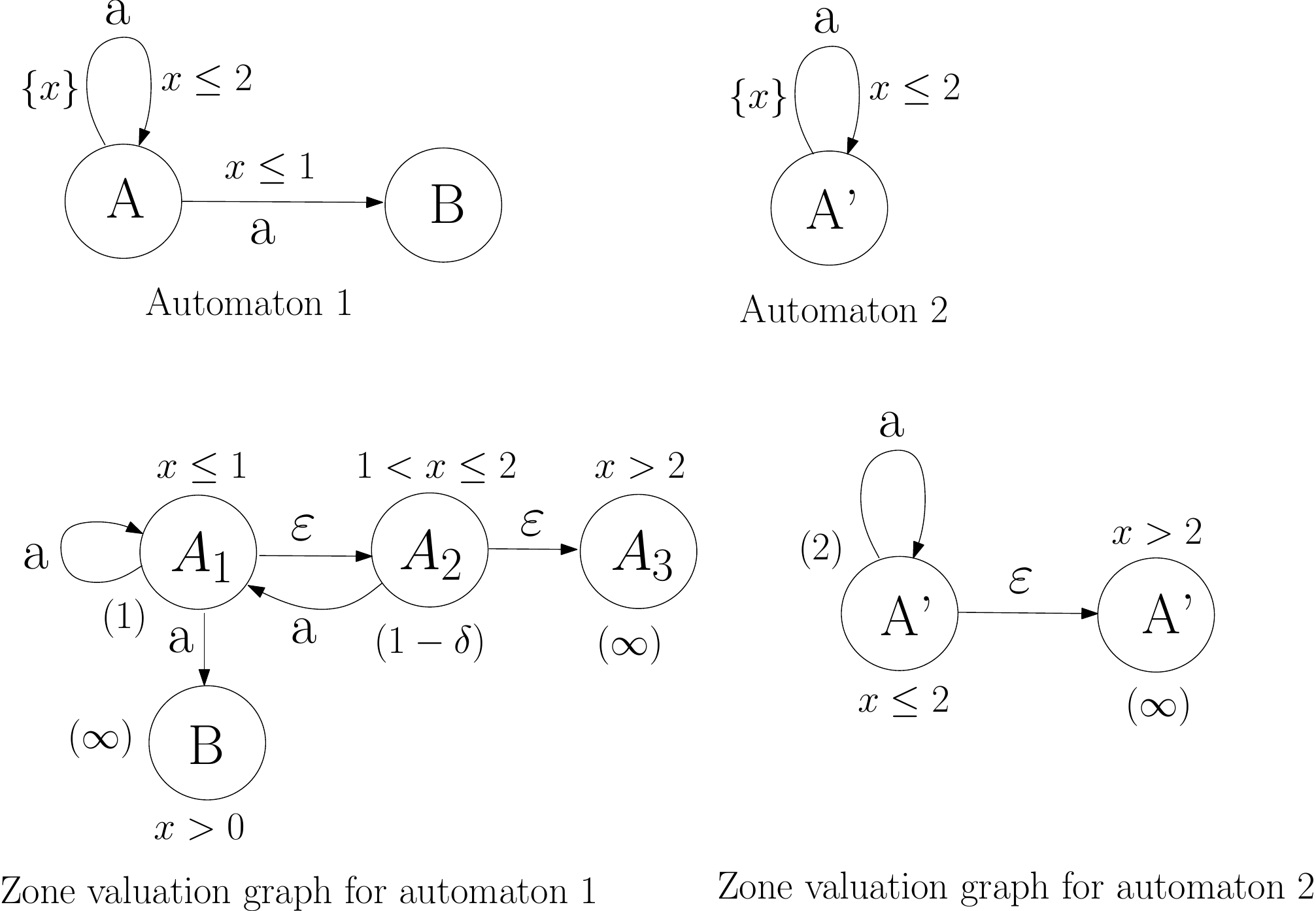}
\caption{\label{fig-exTimedSimEquiv} $0-\Gamma^{Z, \langle a, a \rangle, =}_{\infty}$ game does not characterize timed simulation equivalence. It is characterized by $0-\Gamma^{Z_{sim}, \langle a, a \rangle, =}_{\infty}$.}
\end{figure}	
\end{example}

Phase 2 is modified so as to merge the nodes that are simulation equivalent. Here $A_1$ and $A_2$ are simulation equivalent and thus can be merged to get $Z_{sim}$ on which the game can be played.The nodes of the graph $Z_{sim}$ denote the simulation equivalent classes of the corresponding timed LTS. The defender here has a universal winning strategy when the game is played on this variant of the zone valuation graph.
\begin{lemma}
The game $0-\Gamma_{\infty}^{Z_{sim}, \langle a, a \rangle, =}$ characterizes timed simulation equivalence.
\end{lemma}

\subsection{Timed Performance Prebisimulation Game}
In \cite{GNA1}, it has been shown that two timed processes are timed performance prebisimilar iff their zone valuation graphs are strongly bisimilar and for each pair of strongly bisimilar nodes, all nodes from one zone valuation graph should be equal to or smaller than the corresponding bisimilar node of the other graph. We can design the game as disjunction of two games. In the first game, while checking if the zone valuation graphs $G_1$ and $G_2$ are strongly bisimilar, we also check if the spans of the nodes of graph $G_1$ is less than or equal to the spans of corresponding bisimilar nodes of graph $G_2$. If the defender loses this game, then the second game is played which differs from the first subgame in the extra condition that now it is checked that if the span of the nodes in graph $G_2$ is less than or equal to the span of the bisimilar nodes of $G_1$. The game described above thus is $\Gamma_{\infty}^{Z, \langle a, a\rangle, (G_1, \le)} \: \vee \: \Gamma_{\infty}^{Z, \langle a, a\rangle, (G_2, \le)}$.
\begin{lemma}
The game $\Gamma_{\infty}^{Z, \langle a, a\rangle, (G_1, \le)} \: \vee \: \Gamma_{\infty}^{Z, \langle a, a\rangle, (G_2, \le)}$ characterizes timed performance prebisimulation.
\end{lemma}
\begin{example}
In this example, we consider the two timed automata from \cite{GNA1}. The two timed automata in figure \ref{fig-prebisimexample} are related through timed prebisimulation relation. The automaton in the left is \emph{at least as fast as} the automaton on the right, since the second $a$ action should be performed within a time interval of one time unit after the first $a$ action whereas in the second timed automaton, the second $a$ can be performed within an interval of two time units after the first action. The game $\Gamma_{\infty}^{Z, \langle a, a\rangle, (G_1, \le)} \: \vee \: \Gamma_{\infty}^{Z, \langle a, a\rangle, (G_2, \le)}$ is played on the their corresponding zone valuation graphs which are shown in figure \ref{fig-prebisimexamplecanozone}. The defender has a universal winning strategy.
\begin{figure}[t]
\centering
\includegraphics[width=0.6\textwidth]{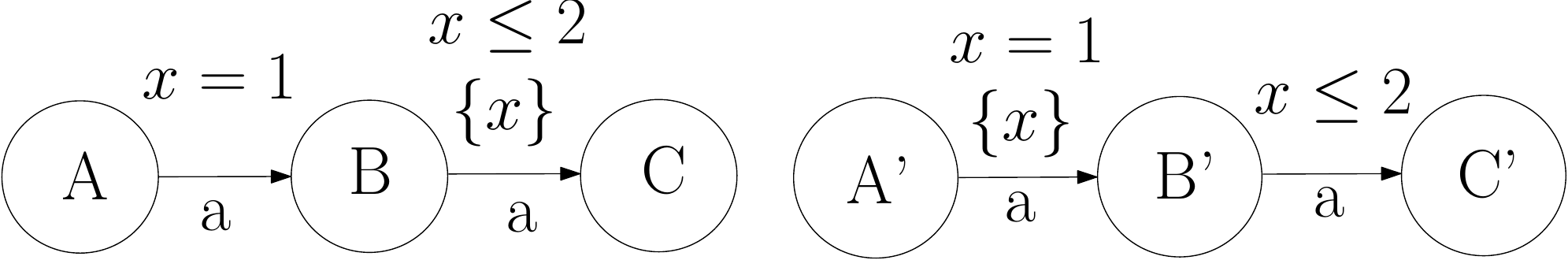}
\caption{\label{fig-prebisimexample} Example: Timed prebisimulation relation}
\end{figure}
\begin{figure}[t]
\centering
\includegraphics[width=0.7\textwidth]{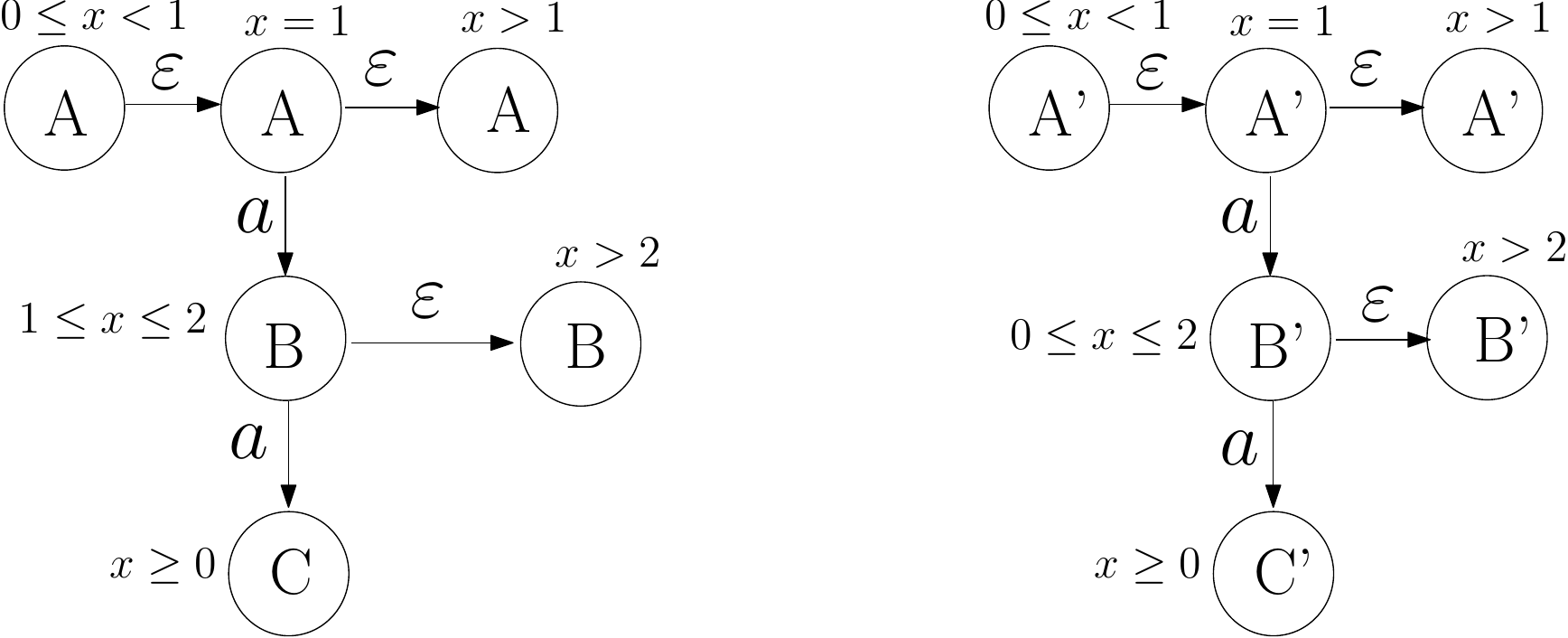}
\caption{\label{fig-prebisimexamplecanozone} Example: Zone valuation graph of timed automata shown in figures \ref{fig-prebisimexample}}
\end{figure}
\end{example}

\section{Hierarchy of Games} \label{sec-gamehier}
The following lemmas describe the hierarchy across different timed games that are obtained by assigning different values to each of the parameters in the game template. The arrow from the game on the left to the game on the right denotes that if the defender has a universal winning strategy for the game on the left, then it also has a universal winning strategy for the game on the right. Besides for each pair of games, if $\Gamma_1 \longrightarrow \Gamma_2$, then $\Gamma_2 \not \longrightarrow \Gamma_1$.
\begin{figure}
\centering
\includegraphics[width=0.9\textwidth]{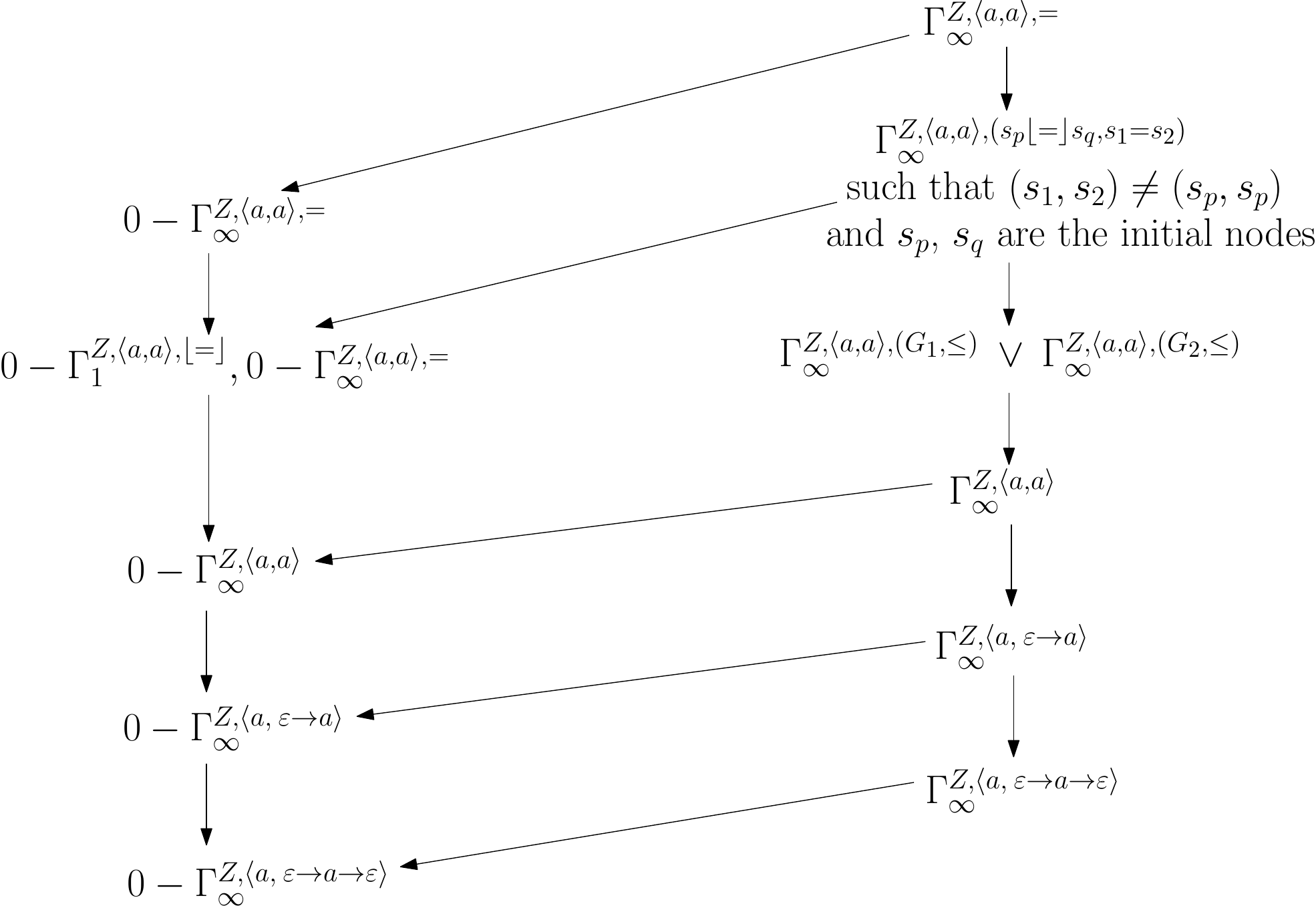}
\caption{\label{fig-gamehierarchy} Hierarchy of timed games}
\end{figure}	
\begin{lemma}
$\Gamma_{\infty}^{G, \alpha, \beta} \longrightarrow n-\Gamma_{\infty}^{G, \alpha, \beta} $
\end{lemma}
This lemma states that if the defender has a universal winning strategy in a game with no restriction on alternations, then it will also win a game with finite number of alternations if the other parameters do not change.
\begin{lemma}
$\Gamma_{\infty}^{G, \alpha, \beta} \longrightarrow \Gamma_{k}^{G, \alpha, \beta} $
\end{lemma}
This lemma states that if the defender has a universal winning strategy in a game with infinite number of rounds, then it will also win in a game with finite number of rounds.
\begin{lemma}\label{lem-third}
$n-\Gamma_{k}^{G, \alpha, =} \longrightarrow n-\Gamma_{k}^{G, \alpha, \lfloor = \rfloor} $\\
$n-\Gamma_{k}^{G, \alpha, =} \longrightarrow n-\Gamma_{k}^{G, \alpha, (G_1, \le)} $\\
$n-\Gamma_{k}^{G, \alpha, =} \longrightarrow n-\Gamma_{k}^{G, \alpha, (G_2, \le)} $\\
$n-\Gamma_{1}^{G, \alpha, \lfloor = \rfloor} \longrightarrow n-\Gamma_{1}^{G, \alpha, (G_1, \le)} \: \vee \: n-\Gamma_{1}^{G, \alpha, (G_2, \le)}$\\
$n-\Gamma_{k}^{G, \alpha, \beta} \longrightarrow n-\Gamma_{k}^{G, \alpha} $
\end{lemma}
\begin{corollary}
$\Gamma^{Z, \langle a, a \rangle, (s_p \lfloor = \rfloor s_q, s_1 = s_2)}_\infty$ such that $(s_1, s_2) \neq (s_p, s_q)$ $\longrightarrow \Gamma_{\infty}^{Z, \langle a, a\rangle, (G_1, \le)} \:\vee \:\Gamma_{\infty}^{Z, \langle a, a\rangle, (G_2, \le)}$
\end{corollary}
This is immediate from lemma \ref{lem-third}.
\begin{lemma}
$n-\Gamma_{k}^{G, \langle a, a\rangle, \beta} \longrightarrow n-\Gamma_{k}^{G, \langle a, \varepsilon \rightarrow a \rangle, \beta} \longrightarrow n-\Gamma_{k}^{G, \langle a, \varepsilon \rightarrow a \rightarrow \varepsilon \rangle, \beta}$
\end{lemma}
This is true since every node in the zone valuation graph has an implicit edge labelled with $\varepsilon$. Here $a \in Lep$.
\begin{lemma}
$n-\Gamma_{k}^{Z, \langle a, a\rangle, \beta} \longrightarrow n-\Gamma_{k}^{Z_{sim}, \langle a, a\rangle, \beta}$
\end{lemma}
Thus assigning different values to each of these parameters $n_i$, $k_i$, $G_i$, $\alpha_i$, $\beta_i$ in the $i$th subgame, we can generate a complete game hierarchy using the lemmas given above. Below we give a diagram which shows the hierarchy of the games that correspond to the timed relations in figure \ref{fig-timedSpectrum}.
The diagram in figure \ref{fig-gamehierarchy} is only a small part of the entire hierarchy of timed games defined in this paper and as in \cite{CD1}, this leaves us with the scope of defining several timed relations or embed existing relations that are not discussed in this paper into this game hierarchy.

\section{Conclusion} \label{sec-conc}
In this paper, we have presented a hierarchy of games that can be played between two timed processes where these processes denote valuations of timed automata. Timed automata is a well studied formalism and the decidability results corresponding to several relations are known with respect to timed automata. The hierarchy among the games reflects the hierarchy among the timed relations. The game hierarchy also allows us to embed several other timed relations that are not discussed in this paper. The closest to our works are \cite{CS1} and \cite{CD1}. Bisimulation games were first introduced in \cite{CS1} and the game was extended in \cite{CD1} where similar EF games have been designed to characterize process equivalences appearing in Van Glabbeek's spectrum \cite{VG1}. As in \cite{CD1}, in our work too we provide a game template from which the entire hierarchy can be generated by assigning different values to the template parameters. However our case is more difficult since we deal with equivalences and preorders that involve real time. The main challenge here lies in designing the graph structure on which a game has to be played. We found that zone valuation graph introduced in \cite{GNA1} and its variants to be appropriate for this purpose.

\bibliographystyle{plain}
\bibliography{TRversion}

\end{document}